# A modular risk concept for complex systems

Dag McGeorge
*Group Research and Development*
*DNV*
Høvik, Norway
0009-0008-1135-4214

Jon Arne Glomsrud
*Group Research and Development*
*DNV*
Trondheim, Norway
0000-0001-7851-648

***Abstract*—This paper motivates the views that for complex systems, risk should be controlled by enforcing constraints in a modular way at different system levels, that the constraints can be expressed as assurance contracts and that acceptable risk mitigation can be demonstrated in assurance case modules. This short paper explains how already existing methodologies can be combined to create a concept for modular risk assessment. The main novelty is the use of so-called contract-based design (CBD) contracts and refinements as risk constraints. This idea is presented here with the objective of receiving feedback from industry and academia.**

## I. INTRODUCTION

The complexity of systems increases with digitalization. An autonomous ferry, for example, relies on data, algorithms and AI teamed with humans, integrated with barriers and other architectural features etc. Such systems may be understood at different levels, where different actors can be responsible for different subsystems and services. Providers are usually responsible for its products; and owners, users or operators are responsible for operations. A risk assessment is typically made during design, making assumptions about the use and operational contexts which are imposed as requirements on the user or operator who is responsible for not exceeding them.

This fits less well with digitalization, especially AI. Systems are updated frequently and the use and operational context change and are difficult to specify accurately. Components may be used for quite different things, like an AI algorithm trained with different data to do different things, a bit like identical nuts and bolts that are used with different functions in different systems. And the responsibility for all these changes are shared.

This calls for a modular approach to managing risk, which fits any actor in the supply chain and system life cycle.

## II. 2 MOTIVATION AND BACKGROUND

An engineered system is created to provide functions that users value [1]. Along with concerns of third-party stakeholders, they represent interests in the system worthy of attention. Floridi [2] provides a theoretical framework for system modelling concerning differing interests, which we will use referring to the interests as system requirements or *guarantees*.

An attempt at outlining a monolithic safety case for an autonomous ferry [3] demonstrated that, although structured arguments are subdivided into distinct links or steps in a chain (or tree) of argumentation, it was not practically possible to build a monolithic assurance case for a claim about the safe navigation because the argument would require too many nodes and too many complex dependencies between them with shared responsibilities distributed between several parties not necessarily interested in disclosing all details to each other. This motivated the work presented here with the aim of receiving feedback from industry and academia.

Contract-based design (CBD) [4] expresses dependencies between subsystems as contracts, with assumptions that each subsystem expects and under which the subsystem can provide *guarantees*. For complex systems, contracts and *refinements*, which specify dependencies on and between subsystems, are combined into a *specification structure* for the system with its subsystems [4].

Operations come with side effects that must be controlled to avoid unacceptable risk. Hence, risk management can be viewed as a control problem [5], which includes layers of corporate and societal governance to protect us from harm [6]. This control perspective on risk can be used to develop safety requirements that, if effective and complied with, implies acceptable risk [7]. A novel view on safety for autonomous systems claims that safety is controlled by satisfaction of constraints, which include traditional risk mitigation, but also societal risks and legal and ethical issues [8]. These works suggest that risk management must enforce effective constraints of various sorts to control hazards related directly to the system of interest and indirectly from other system levels. Such constraints are exactly what a CBD specification structure can express. And it can be used to define modular assurance cases showing that the constraints are met [9][10]. A new generic recommended practice for assurance of AI-enabled systems [11] states “Assurance […] involves constructing an assurance argument consisting of claims […] substantiated by evidence” and requires a modular approach to assurance without specifying how to do it in practice. The modular risk concept presented here aims to provide a practical approach to such modular assurance.

## III. THE NEW MODULAR RISK CONCEPT

### *A. Overview*

Any actor involved in the lifecycle of the system needs to assess risk at the system level that the actor is responsible for (e.g. a ferry, a navigation function, an AI inside it or a higher-level mobility system). To meet this need, the modular risk assessment

1. identifies stakeholder needs and expresses them as system requirements (guarantees);
2. performs system analysis to identify how subsystems contribute towards the system requirements;
3. expresses the responsibilities for the system and its subsystems as CBD contracts and refinements;
4. assesses risk using the CBD contracts and refinements as constraints that must be respected; and
5. creates assurance case modules showing that the constraints are met.

The first two steps are standard in systems engineering. Step 3 is outlined in Section III.B. The main novelty is step 4, where the CBD specification structure is used to modularize the risk assessment, see Section III.C. Section III.D outlines how this enables constructing a modular assurance case.

*B. Modular system analysis*

The modular system analysis aims to establish CBD contracts and refinements. The system considered, $C$, is distinguished from its environment $C_e$ that includes $C$'s subsystems, any higher-level systems that $C$ is part of and the external environment. A set of system requirements or guarantees $G_i$ is input from Step 1. For each $G_i$ the analysis identifies the assumptions $A_{ij}$ made about $C_e$ that are needed for $G_i$ to hold. This results in a set of assurance contracts, $K_i=(A_{ij}, G_i)$, where the contract is satisfied if $G_i$ holds whenever the set of assumptions $A_{ij}$ holds, see [4] for details.

Then each $G_i$ is assessed to identify which subsystems contribute to it and specify their contributions as guarantees allocated to the subsystems. Refinements are specified for the dependencies on and between subsystems [4].

The analysis is repeated recursively to other system levels of interest to produce the specification structure for the system of interest.

A system model that covers emergence of properties and behaviour from interactions within $C$ and with $C_e$ is needed as basis for the system analysis to ensure the relevance and completeness of the resulting specification structure.

*C. Modular risk identification and assessment*

Compliance with the specification structure depends on the implementation $C$ and its interaction with $C_e$. Both compliance and the specifications themselves suffer from uncertainty inherent in $C$ and $C_e$ and due to limited knowledge about them. The uncertainties are sources of risk.

The use of CBD allows to assess risk of $C$ and its dependency on $C_e$ separately from assessments at the other system levels. The CBD specification structure allows further modularization by:

- Picking any contract assigned to $C$ and risk-assess it separately, asking the question: *Is there a way that the guarantee could not hold even if all assumptions hold?*
- Picking any refinement (dependency) and risk-assess it separately, asking the question: *Is there any way that the refinement could not hold, i.e. that the independent guarantee or assumption can hold while at the same time the dependent assumption does not?*

This modular risk assessment can use existing risk methods such as STPA [7] to i) identify hazards of violating the constraints, ii) prioritize risks and iii) identify safety requirements that enforce the constraints.

*D. The modular assurance case*

Assurance activities construct evidence and arguments justifying confidence that the identified safety requirements are met and hence the contracts satisfied and the risks mitigated. The argument and evidence can be represented as a modular assurance case with:

- one module for each contract (guarantee), and
- one module for each refinement (dependency on and between subsystems).

The modular assurance case shall demonstrate the completeness of and compliance with its contracts and refinements. Then the CBD methodology can be used to combine the assurance modules according to [10] to show acceptable system risk. In this way the argument bridges the gap from the supporting evidence to justified confidence in guarantees about stakeholder needs.

## IV. Discussion and Conclusion

With the aim of enhancing risk management of complex systems, a modular risk concept has been outlined.

By taking a model-based systems approach combined with contract-based design and assurance cases, this modular risk concept allows to start risk assessment at a high abstraction level to inform architectural decisions and then drill down into the various subsystems and components. It also allows to start assuring a low-level component e.g. if it is new or considered uncertain. The component can then later be integrated into the system in a bottom-up manner.

Together this enables an iterative assurance work process in sync with iterative system development processes allowing the most critical uncertainties to be addressed first, and confidence progressively built in the system as its details are worked out and assessed. It also helps identifying needed architecture modifications early before implementing them becomes prohibitively expensive.

Furthermore the modular approach can be a key enabler for frequent deployment of system updates and frequent adaptation to changing contexts of use by reusing all unaffected modules without rework. It also allows to address different system variants and operational modes systemically and with minimal rework.